\documentclass[aps,prd,twocolumn,preprintnumbers,floatfix,showpacs,nofootinbib]{revtex4}
\usepackage[dvips]{graphics}
\usepackage{amsmath,amsfonts,bm}
\usepackage{epsfig,bm}
\usepackage{graphicx}
\usepackage{xcolor}
\begin{document}

\preprint{SI-HEP-2019-03, QFET-2019-03}

\author{Oscar Cat\`a}
\affiliation{Theoretische Physik 1, Universit\"at Siegen, Walter-Flex-Stra\ss e 3, D-57068 Siegen, Germany}       

\author{Thomas Mannel}
\affiliation{Theoretische Physik 1, Universit\"at Siegen, Walter-Flex-Stra\ss e 3, D-57068 Siegen, Germany}  

\title{${}$\\
Linking lepton number violation with $B$ anomalies
}
\begin{abstract}
\noindent Hints of violation of lepton flavor universality in semileptonic $B$ decays have prompted a renewed interest in leptoquarks at the low TeV scale. Among the different scenarios suggested, some happen to violate also lepton number, yet not much attention has been paid to the expected size of the associated lepton number violating processes. In this note we examine this issue. We find that there is a single leptoquark scenario compatible with the current size of the anomalies which also violates lepton number. In this scenario (Majorana) neutrino masses are radiatively generated. With the leptoquark parameters extracted from fitting the flavor anomalies, one actually gets the right order of magnitude for neutrino masses. We examine the associated effective field theories both at the electroweak scale and at the hadronic scale and estimate the size of the most relevant lepton number violating processes.   
\end{abstract}

\maketitle


\section{Introduction}\label{sec:I}

\noindent The indications of lepton flavor non-universality recently found in $b\to c l\nu$~\cite{Lees:2013uzd,Aaij:2015yra,Hirose:2016wfn} and $b\to s l^+l^-$~\cite{Aaij:2014ora,Aaij:2015esa,Aaij:2017vbb} transitions have provided a renewed interest in leptoquarks at the low-TeV scale. While each of the observables taken in isolation has a rather modest significance, they are correlated in such a way that a consistent pattern arises and relatively simple models can accommodate the different deviations. 

There have been several attempts to build scenarios with a single new field. However, with the exception of a vectorial $U_1(3,2,\tfrac{2}{3})$ leptoquark, there is at present consensus that the minimal scenario to account for the different anomalies in semileptonic $B$ decays requires the addition of at least two scalar leptoquarks at the low TeV scale (see e.g.~\cite{Angelescu:2018tyl}). These scenarios are typically implemented in such a way that each leptoquark accounts separately for the bulk of the $b\to c l\nu$ or the $b\to s l^+l^-$ anomalies. Precisely because the anomalies are treated essentially separately, several combinations of leptoquarks are possible.

In order to accommodate the $B$ anomalies the focus is on interactions that violate lepton flavor universality. However, the presence of leptoquarks are known to potentially induce lepton and baryon number violation. In practice, given the experimental bounds on proton decay, baryon number violation cannot be induced at the low-TeV scale. It is therefore reasonable to assume that the leptoquark interactions have a well-defined baryon number. Lepton number violation (LNV) has a different status: $\Delta L=2$ transitions can account for neutrino masses if they are Majorana fermions. This raises the question whether there might be a link between the generation of neutrino masses and anomalies in $b\to c l\nu$ and $b\to s l^+l^-$ transitions.   

At the hadronic scale, where the anomalies are measured, one can safely integrate out the leptoquarks and use the language of effective field theories (EFTs). The leading effects to the anomalies will then come from dimension-six $4$-lepton operators coming from single leptoquark exchanges. Since $4$-lepton effective operators conserve $B-L$, it generically follows that dimension-six effects must also conserve lepton number.

However, if one allows for more than one leptoquark, then lepton number violation can be generated by terms in the potential, and its effects will be seen at different orders in the EFT expansion. At the electroweak scale, lepton number violation would manifest itself in the Weinberg operator at $d=5$ and in a number of $d\geq 7$ effective operators. The actual generation of these operators of course depends on the specific quantum numbers of the leptoquarks. 

It is a well-known result that neutrino masses can be radiatively generated in its minimal version by adding two new scalars on top of the Standard Model (SM) particle content~\cite{Babu:1989fg,Ma:1998dn}. This solution involves the down-quark sector only and, since the generation takes place at one loop, the scalar masses can be much lighter than the GUT scale. 

In this letter we point out that the scalars involved in the generation of Majorana neutrino masses account simultaneously for the anomalies in $R_{K^{(*)}}$ and $R_{D^{(*)}}$. Actually, by using the values of leptoquark couplings and masses needed to accommodate the anomalies, one ends up with the right order of magnitude for neutrino masses. This quantitative agreement is nontrivial, since it requires the interplay of both leptoquarks. 

Since neutrino mass generation is realized at the electroweak scale by a $d=5$ operator, it is the most sensitive process to lepton number violation. A richer phenomenology can be reached if one considers $d\geq 7$ operators. We briefly comment on the bounds that the leptoquark scenario would put on lepton number violating processes that could be studied at the LHCb and Belle II. We show that LNV processes are generically well out of reach for current particle accelerators.    
  
\section{The minimal scenario}\label{sec:II}

\noindent There have been numerous proposals to reproduce the observed discrepancies in $R_{D^{(*)}}$ and $R_{K^{(*)}}$. Despite efforts to correlate both anomalies to a single new physics particle (see e.g.~\cite{Bauer:2015knc,Becirevic:2016yqi}), the current consensus is that global data fits favor solutions with at least two leptoquarks. In most of these solutions, each leptoquark carries the bulk of one of the anomalies (charged or neutral current transitions). Although the space of models is rather constrained, there are different viable scenarios with leptoquarks and reducing further the number of possibilities requires additional information.   
 
As opposed to other new physics scenarios, the presence of leptoquarks immediately opens the way for lepton and baryon number violation. Baryon number is constrained by the stringent bounds on proton decay, but lepton number violation could be the origin of neutrino masses, if we assume that they are Majorana particles. 

One could therefore ask whether any leptoquark scenario that accounts for the $B$ anomalies can also account for neutrino mass generation in a successful way. The existence of such a scenario is strongly constrained by quantum numbers, and there is no guarantee a priori that the size of the $B$ anomalies could also generate neutrino masses at the right order of magnitude.    

\begin{figure}[t]
\begin{center}
\includegraphics[width=4.5cm]{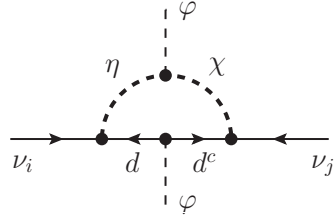}
\end{center}
\caption{\small{\it{Radiative generation of the Weinberg operator.}}}\label{fig:1}
\end{figure}

Since the leptoquark masses hover around the TeV scale, it is clear that quantitative agreement with neutrino masses can only happen if they are radiatively generated. The generation of neutrino masses at one loop was first studied systematically in~\cite{Ma:1998dn}. Among the three different mechanisms that were identified, there is only one which extends the SM with two new particles. The topology is depicted in fig.~\ref{fig:1}, where $d$ stands for a generic down-type quark. 

It is interesting to note that there is no symmetry between the up-type and down-type sector for one-loop neutrino mass generation: a one-loop topology with up-type quarks running into the loop can only happen if additional fields are introduced (see e.g.~\cite{Dorsner:2017wwn}). The down-type sector is therefore naturally selected by the argument of simplicity. 

The quantum numbers of the new scalar particles can then be fixed by going through the diagram of fig.~\ref{fig:1}. This selects the two scalar leptoquark combinations
\begin{align}
\eta\in(3,2,\tfrac{1}{6});\qquad \chi\in({\bar{3}},1,\tfrac{1}{3})
\end{align}
and   
\begin{align}
\eta\in(3,2,\tfrac{1}{6});\qquad \chi_3\in({\bar{3}},3,\tfrac{1}{3})
\end{align}  
Here we will focus on a scenario where the SM is extended just with $\eta$ and $\chi$, which can account for both the $b\to s$ and $b\to c$ anomalies. The scenario with $\eta$ and the electroweak triplet $\chi_3$ is more restricted and will be discussed later on.   
 
Including the leptoquark fields $\eta$ and $\chi$, the SM interactions (in the weak basis) get enlarged by the following operators:
\begin{align}\label{model}
{\cal{L}}&=-\eta^{\dagger}(\Box+m_{\eta}^2)\eta-\chi^*(\Box+m_{\chi}^2)\chi+\mu\chi(\varphi^{\dagger}\eta)\nonumber\\
&-\lambda_1^{ij}{\bar{\ell}}_i\tilde\eta d_j-\lambda_2^{ij}{\bar{q}}^c_{i}\,\epsilon\, \ell_{j}\chi-{\hat{\lambda}}_2^{ij}{\bar{u}}^c_i e_{j}\chi\nonumber\\
&-\lambda_3^{ij}{\bar{q}}^c_{i}\,\epsilon\,q_{j}\chi^*-{\hat{\lambda}}_3^{ij}{\bar{u}}^c_{i}d_{j}\chi^*
\end{align}
where hermitean conjugate operators are implicitly understood and the shorthand notation ${\bar{\psi}}\,\epsilon\,\psi'\equiv{\bar{\psi}}_a\epsilon^{ab}\psi'_{b}$ has been used. The different $\lambda_n^{ij}$ are dimensionless flavor matrices, with $i,j$ generation indices and $a,b$ $SU(2)_L$ indices. 

The previous Lagrangian includes the most general renormalizable interactions of $\chi$ and $\eta$ with SM fields. In eq.~(\ref{model}) we have only omitted quartic terms in the potential, e.g. $(\eta^{\dagger}\eta)(\varphi^{\dagger}\varphi)$ or $\chi^*\chi(\eta^{\dagger}\eta)$, which will not be needed for our discussion.

In order not to violate the bounds on proton decay, we will assume that baryon number is a conserved quantity and choose $B[\eta]=-B[\chi]=\tfrac{1}{3}$. The diquark interactions of the last line will therefore be absent. 

The Lagrangian of eq.~(\ref{model}) can still induce lepton number violating processes whenever the scalar operator $\mu\chi(\varphi^{\dagger}\eta)$ contributes. In other words, the $\mu$-term is a soft breaking of lepton number: when $\mu\to 0$ the leptoquarks decouple from each other and lepton number conservation is restored. This in particular implies that the one-loop diagram of fig.~\ref{fig:1} will generate a finite Weinberg operator. If the $\mu$-term is the only source of lepton number violation, as we will assume, then the interactions of potentially heavier masses cannot affect our results. Notice that, beyond the neutrino mass diagram, the form of the $\mu$-term implies that LNV processes are only induced when {\emph{both}} leptoquarks are involved.  Thus, processes affected by only one of the leptoquarks, regardless of the value of $\mu$, necessarily conserve lepton number (though not necessarily lepton flavor).

The parameter $\mu$ has mass dimensions and is naturally expected to be close to the electroweak scale $v\sim 246$ GeV. More precisely, in order not to upset the Higgs mass at loop order, $\mu\lesssim 4\pi m_h$. In the following we will assume that $\mu\sim v$. 
 
In order to avoid conflict with direct detection at the LHC, the leptoquarks have to be at least around the TeV scale, i.e. heavy as compared to the electroweak scale. This means that they can always be treated as virtual particles and an EFT language, where their effects are integrated out, is very convenient. 

Integrating out the leptoquarks at tree level generates effects at the $d=6$ level. The resulting effective theory reads:
\begin{align}\label{dim6}
{\cal{L}}_{\mathrm{eff}}^{(6)}&=-\frac{\lambda_1^{ij}\lambda_1^{*kn}}{2m_{\eta}^2}{\bar{d}}_n\gamma^{\mu}d_i{\bar{\ell}}_j\gamma_{\mu}\ell_k\nonumber\\
&+\frac{{\hat{\lambda}}_2^{ij}{\hat{\lambda}}_2^{*kn}}{2m_{\chi}^2}{\bar{u}}_n\gamma^{\mu}u_i{\bar{e}}_k\gamma_{\mu}e_j\nonumber\\
&+\frac{\lambda_2^{ij}\lambda_2^{*kn}}{2m_{\chi}^2}\epsilon_{ab}\epsilon_{df}{\bar{q}}^n_a\gamma^{\mu}q^i_d{\bar{\ell}}_b^k\gamma_{\mu}\ell_f^j\nonumber\\
&-\frac{\lambda_2^{ij}{\hat{\lambda}}_2^{*kn}}{2m_{\chi}^2}\epsilon_{ab}\left[{\bar{u}}^nq_a^i{\bar{e}}^k\ell_b^j-\frac{1}{4}{\bar{u}}^n\sigma_{\mu\nu}q_a^i{\bar{e}}^k\sigma^{\mu\nu}\ell_b^j\right]
\end{align}
where Fierz transformations have been performed. The previous operators are written in the weak basis. The rotation to the mass basis gives raise to the following flavor matrices:
\begin{align}
\lambda_1^{\nu d}&=U_{\nu}^T\lambda_1 V_d;\quad \lambda_1^{ed}=U_e^T\lambda_1V_d;\\
\lambda_2^{ue}&=U_u^T\lambda_2U_e;\quad \lambda_2^{d\nu}=U_d^T\lambda_2U_{\nu};\quad {\hat{\lambda}}_2^{ue}=V_u^T{\hat{\lambda}}_2 V_e\nonumber
\end{align}
As expected, the presence of leptoquarks selects directions in flavor space which are not dictated by the CKM matrix. The effective operators in eq.~(\ref{dim6}), when runned down to hadronic scales, provide the leading effects to deviations from the SM values for $R_{D^{(*)}}$~\cite{Sakaki:2013bfa,Freytsis:2015qca,Bauer:2015knc} and $R_{K^{(*)}}$~\cite{Hiller:2014yaa}.

Specifically, the anomaly in $R_{K^{(*)}}$ can be accounted for with the following non-zero matrix entries
\begin{align}
\lambda_1^{ed}=\{\lambda_{\mu s},\lambda_{\mu b}\}
\end{align} 
provided they hover around ${\cal{O}}(10^{-2})$ for a low-TeV leptoquark~\cite{Becirevic:2015asa}. Regarding $R_{D^{(*)}}$, one needs the minimal structure
\begin{align}
\lambda_2^{ue}&=\{\lambda_{c\tau}\}\\
\lambda_2^{d\nu}&=\{\lambda_{s\nu_\tau},\lambda_{b\nu_\tau}\}\\
{\hat{\lambda}}_2^{ue}&=\{{\hat{\lambda}}_{c\tau}\}
\end{align}
The outcome of different papers~\cite{Freytsis:2015qca,Bauer:2015knc} shows that for a low-TeV leptoquark the anomalies can be accommodated and constraints on other flavor processes respected if the relevant matrix entries in the flavor matrices $\lambda_j$ are of ${\cal{O}}(10^{-1})$ for the left-handed couplings and ${\cal{O}}(10^{-2})$ for the right-handed ones.  

\section{Neutrino mass generation}

\noindent One can now calculate explicitly the diagram of fig.~\ref{fig:1}. The result is the Weinberg operator~\cite{Weinberg:1979sa},
\begin{align}
{\cal{L}}_{\mathrm{eff}}^{(5)}&=C_5^{ij}({\bar{\ell}}^c_j{\tilde{\varphi}}^*)({\tilde{\varphi}}^{\dagger}\ell_i)\,,
\end{align} 
where $C_5$ depends on which down-type quark runs inside the loop. For $b$ quarks one finds
\begin{align}
C_5^{ij}=\frac{3}{(4\pi)^2}(\lambda_1^{*ib}\lambda_2^{bj})\frac{\mu \lambda_b}{m_\chi^2-m_\eta^2}\log\frac{m_{\chi}^2}{m_{\eta}^2}
\end{align}
where $\lambda_b$ is the bottom Yukawa. The neutrino mass matrix then takes the form
\begin{align}\label{bquark}
m_{\nu}^{ij}=\frac{3}{(4\pi)^2\sqrt{2}}(\lambda_1^{ib}\lambda_2^{*bj})m_b\frac{\mu v}{m_X^2}
\end{align}
which can be diagonalized with a matrix $U_{\nu}$. In order to simplify our results we have assumed that the leptoquarks have comparable masses, $m_{\chi}\sim m_{\eta}=m_X$. 

The contribution of light quarks is qualitatively different. Because of confinement, it is dominated by nonperturbative physics. On dimensional grounds the result takes the form 
\begin{align}
C_5^{ij}\sim -\sqrt{2}\lambda_1^{is}\lambda_2^{*sj}\frac{\mu\langle {\bar{s}} s\rangle}{vm_\eta^2m_\chi^2}
\end{align} 
The contribution to the neutrino mass matrix thus reads
\begin{align}\label{squark}
\delta m_{\nu}^{ij}\sim -\frac{\lambda_1^{is}\lambda_2^{*sj}}{\sqrt{2}}\frac{\mu v}{m_X^4}\langle {\bar{s}} s\rangle
\end{align}

Given the size of the quark condensate, $\langle{\bar{q}}q\rangle\sim -(250\,{\mathrm{MeV}})^3$, the ratio
\begin{align}
\frac{\delta m_{\nu}}{m_{\nu}}\sim -(4\pi)^2\frac{\langle {\bar{q}}q\rangle}{m_X^2m_b}
\end{align} 
gives a negligible ${\cal{O}}(10^{-5})$ relative correction from light quark exchange. The bulk of the neutrino masses is thus given by $b$ exchange.  

Using that $m_{\nu}\lesssim 0.1$ eV, eq.~(\ref{bquark}) can be written as
\begin{align}
m_{\nu}\sim 10^{-3}\lambda_1^{\nu b}\lambda_2^{*b\nu}\left(\frac{1\,{\mathrm{TeV}}}{m_X}\right)^2{\mathrm{GeV}}\lesssim 10^{-10} {\mathrm{GeV}}
\end{align}
The bound can be saturated with leptoquark masses in the low TeV range and $\lambda_j\sim {\cal{O}}(10^{-2}-10^{-3})$. This is precisely the order of magnitude for masses and couplings needed to reproduce the $B$ anomalies. Notice however that the entries of the flavor matrices needed for neutrino masses are not the same as for $R_{K^{(*)}}$ and $R_{D^{(*)}}$. Their values will be constrained instead by processes like $b\to s{\bar{\nu}}\nu$, which currently have rather loose bounds.  

The successful generation of Majorana neutrino masses with the parameters that reproduce the anomalies in $R_{K^{(*)}}$ and $R_{D^{(*)}}$ allows one to reverse the argument: if one believes that neutrinos are Majorana particles, and that the dynamical origin of its mass is much below the GUT scale, then the simplest scenario is that containing two scalar leptoquarks. If one further assumes that the flavor couplings have only a mild hierarchy, anomalies in both charged and neutral currents in $B$ physics should be generated at the level found experimentally. 
 
\section{Other lepton number violating processes}

\noindent In the previous section we have already discussed that the Weinberg operator is the leading operator which violates lepton number. Phenomenologically, however, this operator is rather limited and accounts basically for neutrino mass generation. It is therefore interesting to explore which other processes induced by the leptoquarks $\eta$ and $\chi$ would violate lepton number and could in principle be detected at LHCb and Belle II. 

\begin{figure}[t]
\begin{center}
\includegraphics[width=4.5cm]{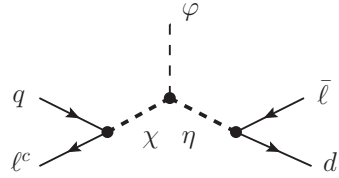}
\end{center}
\caption{\small{\it{Generation of $d=7$ operators through $\eta$ and $\chi$ exchange. Integration of the leptoquarks leads to the operators of eq.~(\ref{dim7}).}}}\label{fig:2}
\end{figure}

Having imposed baryon number conservation, all the $d=6$ operators that can be generated conserve lepton number as well. At the electroweak scale, operators that violate lepton number and conserve baryon number will appear next at $d=7$. For the model we are considering, these effective operators are the result of configurations where both leptoquarks are exchanged. The relevant topology is shown in fig.~\ref{fig:2}, which can be constructed by opening up the loop diagram of fig.~\ref{fig:1} and thus adding two extra external fermions. After integrating out the leptoquarks, the resulting effective operators read
\begin{align}\label{dim7}
{\cal{L}}_{\mathrm{eff}}^{(7)}&=\frac{\mu}{m_{\eta}^2m_{\chi}^2}\lambda_1^{ij}\lambda_2^{*kn}({\bar{\ell}}_i\tilde{\varphi}d_j)({\bar{q}}_k\,\epsilon\,\ell^c_n)\nonumber\\
&+\frac{\mu}{m_{\eta}^2m_{\chi}^2}\lambda_1^{ij}{\hat{\lambda}}_2^{*kn}({\bar{\ell}}_i\tilde{\varphi}d_j)({\bar{u}}_k e_n^c)\,,
\end{align} 
which, as expected, violate lepton number by two units. If we define $\Delta Q_{\ell}$ as the difference of lepton charge, they induce processes with $\Delta L=2$, $\Delta B=0$ and $\Delta Q_{\ell}=0,1$, i.e. $d_i\to d_j \nu\nu$ or $d_i\to u_j l\nu$. 

At the hadronic level, the operators in eq.~(\ref{dim7}) appear as 4-lepton operators, but with a relative suppression factor of
\begin{align}
\frac{\mu v}{m_X^2}
\end{align} 
with respect to those coming from eq.~(\ref{dim6}). On top of this, since the processes they induce do not interfere with the Standard Model ones, their effects will generically be very suppressed. 

Consider for instance the $\Delta L=2$ process $b\to c\tau^-\nu$. Compared with $b\to c\tau^-{\bar{\nu}}$, one gets a correction
\begin{align}
\frac{{\cal{B}}_{b\to c \tau^-\nu}}{{\cal{B}}_{b\to c \tau^-{\bar{\nu}}}}\sim 10^{-3}(\lambda_1^{b\nu}\lambda_2^{*c\tau})^2 \left(\frac{1\, {\mathrm{TeV}}}{m_X}\right)^8
\end{align} 
which optimistically would hover around $10^{-8}$. 

For the corresponding $b\to s$ transitions the ratio is substantially larger, because the process to compare to is loop-suppressed in the Standard Model. Thus,
\begin{align}
\frac{{\cal{B}}_{b\to s\nu\nu}}{{\cal{B}}_{b\to s\nu{\bar{\nu}}}}\sim 10^3(\lambda_1^{b\nu}\lambda_2^{*s\nu})^2\left(\frac{1\, {\mathrm{TeV}}}{m_X}\right)^8
\end{align} 
Since the current experimental bounds are roughly a factor $4$ above the Standard Model prediction, the constraint is satisfied for $(\lambda_1^{b\nu}\lambda_2^{*s\nu})\lesssim 10^{-2}$ with $m_X\simeq 1$ TeV. 

Given that neutrinos are not distinguished from antineutrinos at colliders, the previous estimate indicates that there could be a potentially sizeable lepton number violating contribution to the $b\to s+2\nu$ measurement. However, for the same reason that neutrinos and antineutrinos cannot be separated apart, excesses in $b\to s\nu{\bar{\nu}}$ would point at new physics but could not be ascribed unambiguously to violations of lepton number.  

\begin{figure}[t]
\begin{center}
\includegraphics[width=4.5cm]{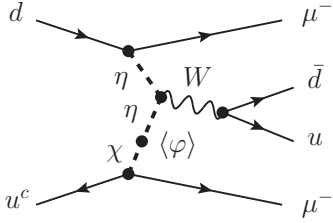}
\end{center}
\caption{\small{\it{Generation of the dominant $\Delta Q=2$ processes through $W$ exchange.}}}\label{fig:3}
\end{figure}

Instead, processes with $\Delta L=2$ and $\Delta Q_{\ell}=2$, i.e. with two charged leptons of the same sign, would leave very distinct signatures and directly test lepton number violation. In the model we are considering, these processes are generated through the diagram of fig.~\ref{fig:3}. Once the leptoquarks are integrated out, one ends up with the following $d=9$ effective operators (up to hermitean conjugation) at the electroweak scale:
\begin{align}\label{dnine}
{\cal{L}}_{\mathrm{eff}}^{(9)}&=\frac{\mu}{m_{\chi}^2m_{\eta}^4}\lambda_1^{ij}\lambda_2^{*kl}D_{\mu}\big[({\bar{\ell}}_i\,\epsilon q^c_j)\varphi^{\dagger}_a\big]D^{\mu}\big[{\bar{d}}^c_k(\epsilon \ell^c_l)^a\big]\nonumber\\
&+\frac{\mu}{m_{\chi}^2m_{\eta}^4}\lambda_1^{ij}{\hat{\lambda}}_2^{*kl}D_{\mu}\big[({\bar{e}}_i\,u^c_j)\varphi^{\dagger}_a\big]D^{\mu}\big[{\bar{d}}^c_k(\epsilon \ell^c_l)^a\big]
\end{align}
In order to induce LNV processes with violation of lepton charge by two units such as the ones in fig.~\ref{fig:3} one needs to pull a $W$ boson out of the covariant derivatives above.

At the electroweak scale, a representative process described by the operators of eq.~(\ref{dnine}) is $t\to bW^-\mu^+\mu^+$. In order to get an approximate order of magnitude, the decay rate can be estimated by factorizing the 4-body phase space and assuming that the final-state particles are massless. One then finds that 
\begin{align}
\Gamma[t\to bW^-\mu^+\mu^+]
\sim 10^{-14}(\lambda_1\lambda_2)^2 \left(\frac{1\, {\mathrm{TeV}}}{m_X}\right)^{12}{\mathrm{GeV}}
\end{align} 
Since this process is background-free, it would require at least $10^{16}$ top decays for detection, a number which is out of the capabilities of current detectors.   

Other representative processes concern $B$ decays. In this case, it is convenient to work with operators defined at the hadronic scale, obtained by integrating the $W$ and Higgs fields. Neglecting effects associated with the running of scales, the leading contribution comes as a $d=10$ operator, which takes the form
\begin{align}
\frac{4G_Fv\mu}{m_{\chi}^2m_{\eta}^4}\lambda_1^{ij}{\lambda}_2^{*kl}V_{np}\big[\partial_{\mu}({\bar{e}}_iu^c_j)({\bar{d}}^c_ke^c_l)\big]{\bar{u}}_n\gamma^{\mu}d_p
\end{align}
To get an estimate of how big $\Delta Q_{\ell}=2$ effects in $B$ decays can be, consider $B^-\to\pi^+\mu^-\mu^-$. Compared to the related lepton-flavour-conserving process $B^-\to\pi^-\mu^+\mu^-$, one finds
\begin{align}
\frac{{\cal{B}}_{B^-\to\pi^+\mu^-\mu^-}}{{\cal{B}}_{B^-\to\pi^-\mu^+\mu^-}}\simeq(\lambda_1\lambda_2)^2\left(\frac{1\, {\mathrm{TeV}}}{m_X}\right)^{12}10^{-18}
\end{align} 
Since experimentally ${\cal{B}}_{B^-\to\pi^-\mu^+\mu^-}\sim 10^{-8}$, one expects the branching ratio for the LNV process at $10^{-26}$, clearly out of Belle II reach. 

Hadronic $\tau$ decays like $\tau^-\to \pi^+e^-e^-\nu$ or $\tau^+\to \mu^-\pi^+\pi^+$ are also generated at $d=10$. For the former decay mode there are no limits available. The latter has a current experimental bound at $10^{-8}$. In both cases the effects predicted by the leptoquark scenario considered in this paper are unobservable.

The previous estimates for $\Delta Q_{\ell}=2$ processes are based on a specific leptoquark construction. However, one can show that the statements about their undetectability are rather generic by using an EFT argument. At the electroweak scale, $\Delta Q_l=\pm 2$ operators first appear at $d=7$, and are restricted to~\cite{Lehman:2014jma}
\begin{align}\label{d7}
({\bar{d}}\gamma^{\mu}u){\bar{\ell}}^c\,\epsilon\,(D_{\mu} \ell)\quad {\mathrm{and}}\quad ({\bar{d}}\gamma^{\mu}u){\bar{\ell}}^c\,\epsilon\,\sigma_{\mu\nu}(D^{\nu} \ell)
\end{align}
Based on dimensional grounds, these operators are weighted by a coefficient that scales like $\Lambda^{-3}$, where $\Lambda$ is the new-physics scale. However, it is easy to realize that such operators cannot be generated from the tree level exchange of a heavy particle. Given a UV model, these operators will, at the most, be generated at the one-loop level. Therefore, at the hadronic scale, the $\Delta Q_{\ell}=2$ part of, e.g., the first operator in eq.~(\ref{d7}) will be of the form
\begin{align}
\frac{C_{ijklpr}}{(4\pi)^2}\frac{G_F}{\Lambda^3} ({\bar{d}}_i\gamma_{\mu} u_j)({\bar{d}}_k\gamma^{\mu} u_l) ({\bar{e}}^c_p e_r) 
\end{align}
where $\Lambda$ should be understood as the geometric mean of the new-physics masses and $C_{ijklpr}$ is a flavor matrix. Assuming that $\Lambda~\sim {\cal{O}}$ (few TeV), one could, e.g., enhance the decay rate for {\mbox{$B^-\to\pi^+\mu^-\mu^-$}} to
\begin{align}
\frac{{\cal{B}}_{B^-\to\pi^+\mu^-\mu^-}}{{\cal{B}}_{B^-\to\pi^-\mu^+\mu^-}}\lesssim 10^{-12}    
\end{align}
This is an optimistic upper bound. A model with all new particles at the low TeV scale is hardly realistic. In practice, compliance with flavor constraints will push some of the masses up and thus lower this ratio. $10^{-12}$ is therefore to be understood as a generous upper bound, which is clearly too suppressed to be detected at $B$ factories. 


\section{Scenarios with a leptoquark triplet}

\noindent The criteria we used to build our model was its ability to generate neutrino masses. We have thus focussed our attention on the leptoquark pair $\eta(3,2,\tfrac{1}{6})$ and $\chi({\bar{3}},1,\tfrac{1}{3})$. However, we already pointed out that, regarding neutrino mass generation, an alternative scenario would be to consider $\eta(3,2,\tfrac{1}{6})$ and $\chi_3({\bar{3}},3,\tfrac{1}{3})$. It is instructive to highlight the differences between both scenarios for the phenomenological applications discussed in this paper.     

Imposing baryon number conservation, there is only one operator coupling $\chi_{3}$ to fermions, and the Lagrangian reads 
\begin{align}\label{triplet}
{\cal{L}}&=-\eta^{\dagger}(\Box+m_{\eta}^2)\eta-\chi^{\dagger}_3(\Box+m_{\chi}^2)\chi_3+\mu(\varphi^{\dagger}\chi_3\eta)\nonumber\\
&-\lambda_1^{ij}{\bar{\ell}}_i\tilde\eta d_j-\lambda_2^{ij}{\bar{q}}^c_{i}i\tau_2\chi_3 \ell_{j}
\end{align}

Below the TeV scale the leptoquarks can be integrated out. The leading lepton number conserving processes appear at the $d=6$ level. Focusing on $\chi_3$ exchange, one obtains the following effective operators: 
\begin{align}
{\cal{L}}_{eff}^{(6)}=\frac{\lambda_{2}^{ij}\lambda_2^{*kn}}{m_{\chi}^2}\left[({\bar{q}}^c_i \epsilon \ell_j)({\bar{\ell}}_k\epsilon q^c_n)-({\bar{q}}_n\gamma_{\mu} q_i)({\bar{\ell}}_k\gamma^{\mu}\ell_j)\right]
\end{align} 
From the previous operators it is clear that $\chi_3$ contributes to $b\to s$ transitions~\cite{Alonso:2015sja}, and not significantly to $b\to c$. The phenomenology of this operator at the hadronic scale has actually been studied in~\cite{Becirevic:2018afm} and shown to be able to accommodate the $R_{K^{(*)}}$ anomaly. 

This means that a scenario with $\eta(3,2,\tfrac{1}{6})$ and $\chi_3({\bar{3}},3,\tfrac{1}{3})$ will generate neutrino masses but can only account for discrepancies in the neutral $b\to s$ transitions. This scenario has already been investigated in~\cite{Pas:2015hca,Cheung:2016fjo}, where the motivation for having these two leptoquarks was to accommodate neutrino masses and $R_K$ simultaneously. 

The advantage of a scenario with $\chi({\bar{3}},1,\tfrac{1}{3})$ and $\eta(3,2,\tfrac{1}{6})$ is that it is far more encompassing: one can describe simultaneously the anomalies in $b\to c$ and $b\to s$ transitions and at the same time have a mechanism for neutrino mass generation. Additionally, it has been shown that $\chi({\bar{3}},1,\tfrac{1}{3})$ can also generate an effect on the muon anomalous magnetic moment~\cite{Queiroz:2014zfa} compatible with the discrepancy observed experimentally~\cite{ColuccioLeskow:2016dox}.  

Regarding lepton number violating processes, there will be qualitatively no changes if $\chi$ is replaced by $\chi_3$. Neutrino masses will be generated by the $t_3=0$ component of the triplet and the formulas given above will get modified trivially with the appropriate replacements of the $\lambda$ matrices. The lepton number violating processes depicted in fig.~\ref{fig:2} will be generated with two of the components of $\chi_3$. This will add an extra diagram, but will generate the same effective operators at the hadronic scale. Similarly, for $\Delta Q_{\ell}=2$ processes, there will also be an extra topology to the one of fig.~\ref{fig:3}, with the $W$ exchanged between the different components of $\chi_3$. The bounds that we found would only get affected by ${\cal{O}}(1)$ effects.


\section{Conclusions}

\noindent The tensions observed in semileptonic $B$ decays, if confirmed, would be a clear signal of lepton universality violation. However, the fact that the most natural scenarios to accommodate the discrepancies involve leptoquarks also suggests that lepton number could be violated.  

In this paper we have entertained this idea and explored its consequences. With only one leptoquark, lepton number violation would imply baryon number violation. With two leptoquarks, which is what data currently seems to favor (among the one-leptoquark solutions, only the $U_1(3,2,\tfrac{2}{3})$ model is not excluded), one can accommodate neutrino masses while having a stable proton. We have only considered minimal extensions of the Standard Model that can account for the anomalies in both $b\to s$ and $b\to c$ transitions, namely scenarios with two scalar leptoquarks. Interestingly, there is a single scenario that can explain $R_K^{(*)}$ and $R_D^{(*)}$ and also violate lepton number, with the following set of leptoquarks:
\begin{align}
\eta(3,2,\tfrac{1}{6});\qquad \chi({\bar{3}},1,\tfrac{1}{3})
\end{align}
The dominant lepton number violating effect is the generation of Majorana neutrino masses at the one-loop level. It is remarkable that the present size of the deviation in $R_{K^{(*)}}$ and $R_{D^{(*)}}$ give the right order of magnitude for neutrino masses. Since $\chi({\bar{3}},1,\tfrac{1}{3})$ also bears a contribution to the muon anomalous magnetic moment able to explain the current discrepancy with the Standard Model prediction, this scenario is rather attractive. 

Beyond neutrino masses, the effects of the leptoquarks on lepton number violation processes are extremely suppressed, currently at an undetectable level. One could have ${\cal{O}}(1)$ deviations in $b\to s \nu\nu$ transitions, but since neutrinos and antineutrinos cannot be distinguished at the detector level, this potential deviation of Standard Model physics would not be a conclusive signal of lepton number violation.   

If the deviations in the $b\to s$ and $b\to c$ transitions persist and this model is taken seriously, it would then indicate that with the $B$ anomalies we are actually probing the scale of lepton number violation, which would also be the scale of flavor lepton universality breaking and would turn out to be at the low-TeV scale. 

\section*{Acknowledgements}
We thank Thorsten Feldmann for reading the manuscript and for very stimulating discussions. This work is supported in part by the Deutsche Forschungsgemeinschaft (DFG FOR 1873).


\begin{thebibliography}{999}

\bibitem{Lees:2013uzd} 
  J.~P.~Lees {\it et al.} [BaBar Collaboration],
  Phys.\ Rev.\ D {\bf 88}, no. 7, 072012 (2013)
  [arXiv:1303.0571 [hep-ex]].
  
\bibitem{Aaij:2015yra} 
  R.~Aaij {\it et al.} [LHCb Collaboration],
  Phys.\ Rev.\ Lett.\  {\bf 115}, no. 11, 111803 (2015)
  Erratum: [Phys.\ Rev.\ Lett.\  {\bf 115}, no. 15, 159901 (2015)]
  [arXiv:1506.08614 [hep-ex]].
	
\bibitem{Hirose:2016wfn} 
  S.~Hirose {\it et al.} [Belle Collaboration],
  Phys.\ Rev.\ Lett.\  {\bf 118}, no. 21, 211801 (2017)
  [arXiv:1612.00529 [hep-ex]].

\bibitem{Aaij:2014ora} 
  R.~Aaij {\it et al.} [LHCb Collaboration],
  Phys.\ Rev.\ Lett.\  {\bf 113}, 151601 (2014)
  [arXiv:1406.6482 [hep-ex]].

\bibitem{Aaij:2015esa} 
  R.~Aaij {\it et al.} [LHCb Collaboration],
  JHEP {\bf 1509}, 179 (2015)
  [arXiv:1506.08777 [hep-ex]].
	
\bibitem{Aaij:2017vbb} 
  R.~Aaij {\it et al.} [LHCb Collaboration],
  JHEP {\bf 1708}, 055 (2017)
  [arXiv:1705.05802 [hep-ex]].

\bibitem{Angelescu:2018tyl} 
  A.~Angelescu, D.~Be\v cirevi\'c, D.~A.~Faroughy and O.~Sumensari,
  JHEP {\bf 1810}, 183 (2018)
  [arXiv:1808.08179 [hep-ph]].
		
\bibitem{Babu:1989fg} 
  K.~S.~Babu and E.~Ma,
  Mod.\ Phys.\ Lett.\ A {\bf 4}, 1975 (1989).
		
\bibitem{Ma:1998dn} 
  E.~Ma,
  Phys.\ Rev.\ Lett.\  {\bf 81}, 1171 (1998)
  [hep-ph/9805219].

\bibitem{Bauer:2015knc} 
  M.~Bauer and M.~Neubert,
  Phys.\ Rev.\ Lett.\  {\bf 116}, no. 14, 141802 (2016)
  [arXiv:1511.01900 [hep-ph]].

\bibitem{Becirevic:2016yqi} 
  D.~Be\v cirevi\'c, S.~Fajfer, N.~Ko\v snik and O.~Sumensari,
  Phys.\ Rev.\ D {\bf 94}, no. 11, 115021 (2016)
  [arXiv:1608.08501 [hep-ph]].
		
\bibitem{Dorsner:2017wwn} 
  I.~Dor\v sner, S.~Fajfer and N.~Ko\v snik,
  Eur.\ Phys.\ J.\ C {\bf 77}, no. 6, 417 (2017)
  [arXiv:1701.08322 [hep-ph]].
	
\bibitem{Sakaki:2013bfa} 
  Y.~Sakaki, M.~Tanaka, A.~Tayduganov and R.~Watanabe,
  Phys.\ Rev.\ D {\bf 88}, no. 9, 094012 (2013)
  [arXiv:1309.0301 [hep-ph]].

\bibitem{Freytsis:2015qca} 
  M.~Freytsis, Z.~Ligeti and J.~T.~Ruderman,
  Phys.\ Rev.\ D {\bf 92}, no. 5, 054018 (2015)
  [arXiv:1506.08896 [hep-ph]].

\bibitem{Hiller:2014yaa} 
  G.~Hiller and M.~Schmaltz,
  Phys.\ Rev.\ D {\bf 90}, 054014 (2014)
  [arXiv:1408.1627 [hep-ph]].

\bibitem{Becirevic:2015asa} 
  D.~Be?irevi?, S.~Fajfer and N.~Košnik,
  Phys.\ Rev.\ D {\bf 92}, no. 1, 014016 (2015)
  [arXiv:1503.09024 [hep-ph]].

\bibitem{Weinberg:1979sa} 
  S.~Weinberg,
  Phys.\ Rev.\ Lett.\  {\bf 43}, 1566 (1979).

\bibitem{Lehman:2014jma} 
  L.~Lehman,
  Phys.\ Rev.\ D {\bf 90}, no. 12, 125023 (2014)
  [arXiv:1410.4193 [hep-ph]].

\bibitem{Alonso:2015sja} 
  R.~Alonso, B.~Grinstein and J.~Martin Camalich,
  JHEP {\bf 1510}, 184 (2015)
  [arXiv:1505.05164 [hep-ph]].

\bibitem{Becirevic:2018afm} 
  D.~Be\v cirevi\'c, I.~Dor\v sner, S.~Fajfer, N.~Ko\v snik, D.~A.~Faroughy and O.~Sumensari,
  Phys.\ Rev.\ D {\bf 98}, no. 5, 055003 (2018)
  [arXiv:1806.05689 [hep-ph]].
	
\bibitem{Pas:2015hca} 
  H.~P\"as and E.~Schumacher,
  Phys.\ Rev.\ D {\bf 92}, no. 11, 114025 (2015)
  [arXiv:1510.08757 [hep-ph]].

\bibitem{Cheung:2016fjo} 
  K.~Cheung, T.~Nomura and H.~Okada,
  Phys.\ Rev.\ D {\bf 94}, no. 11, 115024 (2016)
  [arXiv:1610.02322 [hep-ph]].

\bibitem{Queiroz:2014zfa} 
  F.~S.~Queiroz and W.~Shepherd,
  Phys.\ Rev.\ D {\bf 89}, no. 9, 095024 (2014)
  [arXiv:1403.2309 [hep-ph]].
  
\bibitem{ColuccioLeskow:2016dox} 
  E.~Coluccio Leskow, G.~D'Ambrosio, A.~Crivellin and D.~Müller,
  Phys.\ Rev.\ D {\bf 95}, no. 5, 055018 (2017)
  [arXiv:1612.06858 [hep-ph]].
			
\end{thebibliography}
\end{document}